\documentclass[a4paper,dvipdfmx,fleqn,usenatbib]{mnras}

\usepackage{newtxtext,newtxmath}

\usepackage[T1]{fontenc}
\usepackage{ae,aecompl}

\usepackage{graphicx}	
\usepackage{amsmath}	
\usepackage{amssymb}	

\title[Donors in NS-ULX]{Nature of donors in ultra-luminous X-ray binaries powered by neutron stars}

\author[S. Karino]{
Shigeyuki Karino$^{1}$\thanks{E-mail: karino@ip.kyusan-u.ac.jp}
\\
$^{1}$Faculty of Science and Engineering, Kyushu Sangyo University, 2-3-1 Matsukadai, 
Fukuoka 813-8503, Japan \\
}

\date{Accepted XXX. Received YYY; in original form ZZZ}

\pubyear{2017}

\begin{document}
\label{firstpage}
\pagerange{\pageref{firstpage}--\pageref{lastpage}}
\maketitle

\begin{abstract}

In this study, we examine the properties of donor stars of the three recently discovered ultraluminous X-ray sources (ULXs) powered by rotating neutron stars. 
For this purpose, we constructed a theoretical relation between the X-ray luminosity ($L_{\rm{X}}$) and the orbital period ($P_{\rm{orb}}$) suitable for ULXs with neutron stars. 
By using this new $L_{\rm{X}} - P_{\rm{orb}}$ relation, we attempt to determine the currently unknown nature of donor stars in ULXs associated with neutron stars. 
Especially, comparing the observed properties with the stellar evolution tracks, we suggest that the donor star in the NGC5907 ULX-1 system is a moderately massive star with $6 - 10 \rm{M}_{\odot}$, just departing from the main sequence phase.
The results of our models for the other two ULX systems (M82 X-2 and NGC7793 P-13) are consistent with those in previous studies. 
Although there are only a few samples, observed ULX systems with neutron stars seems to involve relatively massive donors. 

\end{abstract}

\begin{keywords}
accretion, accretion disk -- stars: neutron -- X-rays: binaries 
\end{keywords}



\section{Introduction}

Ultraluminous X-ray sources (ULX) are bright extragalactic X-ray point sources with observed fluxes that would correspond to luminosities greater than the Eddington limit for normal stellar-mass compact objects, if they were radiating isotropically.
Because of these large luminosities, it has been widely believed that they are associated with intermediate-mass black holes \citep{M00, F09, P14} or stellar-mass black holes accreting at a supercritical rate \citep{B02, OM07}.

Recently, periodic pulsations from some ULXs have been recorded \citep{B14, F16, I17a, I17b}. 
These X-ray pulsations strongly indicate the involvement of rotating magnetized neutron stars as the power sources of several ULXs, rather than black holes.
However, the peak fluxes corresponding to $L_{X} > 10^{39}\, \rm{erg \, s}^{-1}$ requires adequate explanations of the accretion and emission mechanisms onto neutron stars. 
The true nature of ULXs is still under discussion, although the existence of an extremely strong magnetic field above the quantum limit \citep{E14} and/or the non-spherical accretion and beam emission are suggested to increase the luminosities significantly above the Eddington limit \citep[see, for example][]{OM11}.

The first ULX involving a neutron star (NS-ULX) was found by the NuSTAR satellite. 
\citet{B14} reported NuSTAR observations of ULX M82 X-2 (also known as NuSTAR J095551+6940.8), which revealed periodic changes in the hard X-ray luminosity, indicative of a rotating magnetised neutron star being involved, rather than a black hole. 
The period of the pulsation was found to be $P_{\rm{s}} = 1.37 \, \rm{s}$. 
Additionally, it showed a 2.53-day sinusoidal modulation, interpreted as an orbital period $P_{\rm{orb}}$ corresponding to the motion around an unseen companion, which could be the mass donor in the accreting system. 
The mass donor in this system appears to have a mass greater than $5.2 \rm{M}_{\odot}$.
Though the maximum luminosity reaches 
$ \sim 10^{40} \rm{erg \, s}^{-1}$, 
it becomes an undetectably dim source in the quiescent phase \citep{DSD14}. 
It is suggested that this transient behaviour is caused by a propeller effect, due to a strong magnetic field preventing material accretion \citep{T16}.

In 2016, two more findings of NS-ULXs have been reported \citep{F16, I17a, I17b}. 
One is NGC5907 ULX-1 (hereafter, ULX-1), which shows a 1.1 s pulsation, and another is NGC7793 ULX P-13 (hereafter, P-13) with a 0.4 s pulse period. 
The orbital period was found to be 5.3 days for ULX-1 \citep{I17a} and $64$ days for P-13 \citep{M14}. 
From the optical observations, the donor of P-13 was determined to be a BI9 star with a mass of $18 - 23 \rm{M}_{\odot}$, and this system could have a highly eccentric orbit \citep{M14}. 
However, the mass donor of ULX-1 is unknown; \citet{I17a} argued that both a low-mass donor or a high-mass donor could be possible.

At least two of the newly found NS-ULXs have relatively high-mass ($ > 5 \rm{M}_{\odot}$) donors.
Hence, they are categorized as high-mass X-ray binaries (HMXB) in the traditional classification \citep{C84, B97}.
In most HMXBs, mass transfers proceed via a wind capture process.
However, this type of wind-fed accretion usually cannot achieve a large accretion rate, exceeding the Eddington limit. 
Instead, it is widely believed that NS-ULXs are fed via the Roche lobe overflow (RLOF) mass transfer from the donor \citep{B14}, though also accretion from Be-disc could be considered \citep{KM16}. 
Accretion mechanisms via RLOF are similar to low-mass X-ray binaries (LMXBs), and are quite rare in HMXB systems. 
Although we have only limited examples of NS-ULXs at present, they support the idea that NS-ULXs have intermediate properties between HMXBs and LMXBs.
If this is true, the conventional classification based on donor mass may become out-dated. 
Thus, the true nature of donors in NS-ULX systems attracts our interest.

In this study, we attempt to draw information about donors in ULX systems from limited observational data only. 
For this purpose, we constructed a theoretical relation between the X-ray luminosities (mass accretion rates) and the orbital periods of NS-ULX systems. 
Then, we try to predict properties of the unseen donors in NS-ULXs.
Especially, we discuss the mass of ULX-1; from our estimation, the donor could be a star just starting the Hertzsprung-gap phase, which has a possible mass is of around $10 \rm{M}_{\odot}$. 
In the next section, we discuss the accretion mode of NS-ULX systems and introduce our model.
In section three, we present our results, especially, the application of our model to the unseen donor of ULX-1.
In section four, we discuss the applicability and limits of our model. 
The last section is devoted to conclusions. 


\section{Roche Lobe Overflow Accretion and Luminosity-Orbital Period Relation}

Considering the mass transfer mechanisms in binary systems, three transfer modes might occur: (i) via spherical wind (as for O-type HMXBs), (ii) via disc-shaped wind (as for Be-type HMXBs) or (iii) by Roche-lobe overflow (as for LMXBs) \citep{B97}. 
Since wind-fed accretion cannot cause super-Eddington X-ray luminosity, in NS-ULX systems, RLOF mass transfer would proceed, as is the case of LMXB systems \citep{B14, E14, DPS14, KM16}. 
In tight binary systems, when the donor expands (or when the angular momentum is lost via magnetic stellar wind and gravitational wave radiation), the donor star fills its Roche lobe. 
Then, the envelope of the donor flows into the accretor via the first Lagrangian point. 
The approximated Roche lobe radius is given by the following formula:
\begin{equation}
R_{\rm{RL}} = \frac{0.49 q^{2/3}}{0.6 q^{2/3} + \ln \left(1+ q^{1/3} \right) } a,
\label{eq:Rrl}
\end{equation}
where $q = M_{\rm{D}} / M_{\rm{NS}}$ is the mass ratio between the donor and the accreting neutron star \citep{E83}.
Here, $a$ is the orbital semi-major axis, 
\begin{equation}
a = \left[ \frac{P_{\rm{orb}}^2}{4 \pi^2} G \left( M_{\rm{D}} + M_{\rm{NS}} \right) \right] ^{1/3} ,
\label{eq:a}
\end{equation}
where $G$ denotes the gravitational constant and $P_{\rm{orb}}$ is the orbital period. 
In this study, for simplicity, we assume a circular orbit. 
The evolutionary stage, when the donor fills its Roche lobe depends on the donor mass and the binary separation. 
If the orbital period is known and the neutron star is assumed to have a canonical mass of ($1.4 \rm{M}_{\odot}$), the Roche lobe radius becomes a function of the donor mass only.

The accretion rate via RLOF depends on the evolutionary stage of the donor star.
When the donor star is in the main sequence or in the early Hertzsprung gap stage, the donor has a radiative envelope, and the mass transfer proceeds in thermal timescale, 
\begin{eqnarray} 
\tau_{\rm{th}} &\sim& \frac{GM^2}{RL} \nonumber \\
&=& 3.13 \times 10^{7} \left( \frac{M}{\rm{M}_{\odot}} \right)^{2}
\left( \frac{R}{\rm{R}_{\odot}} \right)^{-1} \left( \frac{L}{\rm{L}_{\odot}} \right)^{-1} \rm{yr} 
\label{eq:tauth} 
\end{eqnarray} 
\citep{KW96}. 
The mass transfer rate from the donor estimated by this time scale becomes
\begin{eqnarray}
\dot{M}_{\rm{D}} &\sim& \frac{M}{\tau_{\rm{th}}} = \frac{RL}{GM} \nonumber \\
&=& 2.01 \times 10^{18} \left( \frac{M}{\rm{M}_{\odot}} \right)^{-1}
\left( \frac{R}{\rm{R}_{\odot}} \right) \left( \frac{L}{\rm{L}_{\odot}} \right) \rm{g \, s}^{-1}.
\end{eqnarray}
%
Since, the above transfer rate is that estimated around the donor, we have to estimate the mass accretion rate at inner region of the accretion disc. 
The mass accretion rate near the neutron star is reduced due to mass-loss via disc outflow.
At the inner region of the accretion disc, the accretion rate could be written as   
\begin{equation}
\dot{M}_{\rm{NS}} = \frac{R_{\rm{in}}}{R_{\rm{out}}} \dot{M}_{D} ,
\end{equation}
where, $R_{\rm{in}}$ and $R_{\rm{out}}$ are the inner and outer radii of the outflowing disc \citep{K17}.
Though we do not know the outer radius of the disc, the result of numerical MHD simulation suggests that the disc is confined in relatively narrow area: $R_{\rm{in}} / R_{\rm{out}} \approx 3/8$ \citep{TO17}. 
Though this result is obtained for a low magnetic field case, this ratio between the inner and outer radii might provide a helpful perspective. 
Taking this relative size of the disc, the estimated mass accretion rate could become roughly $3/8$ smaller than the mass transfer rate from the donor.

Assuming that the potential energy of this transferred matter is converted into X-rays at the neutron star surface, the corresponding luminosity becomes
\begin{eqnarray}
L_{\rm{X}} &\sim& \epsilon \frac{G M_{\rm{NS}} \dot{M}_{\rm{NS}}}{R_{\rm{NS}}}  \nonumber \\
&=& 3.74 \times 10^{38} \epsilon
\left( \frac{M}{\rm{M}_{\odot}} \right)^{-1}
\left( \frac{R}{\rm{R}_{\odot}} \right) \left( \frac{L}{\rm{L}_{\odot}} \right) \rm{erg \, s}^{-1}, 
\label{eq:lx}
\end{eqnarray}
where $\epsilon$ 
is the energy conversion efficiency. 
This efficiency factor is fixed as 0.1 in this study. 
Since $R \propto M^{1/2}$ and $L \propto M^{5}$ for intermediate mass stars, the typical X-ray luminosity given by eq.~(\ref{eq:lx}) can easily become super-Eddington for $M_{\rm{D}} > 2 \rm{M}_{\odot}$.

However, if the donor star is in a giant phase, which has a large convective envelope, the mass transfer proceeds within a dynamical timescale. 
In this case, the mass transfer rate would become very large, 
and the mass would overflow from the Roche lobe. 
Thus, the system would establish a common envelope \citep{R76, MM79}. 
Such a system can no longer be observed as an ULX; rather, it can be observed in infrared regions \citep{I13}.
In the following discussion, we mainly consider the RLOF mass transfer before the donor evolves into a giant.

\bigskip

It is argued that the X-ray luminosities of the persistent LMXBs depend on their orbital periods (in unit of hours); 
\begin{equation}
L_{\rm{X}} \propto P_{\rm{orb}}^{2.5} 
\label{eq:LxPorb}
\end{equation}
\citep{T83}.
This relation was confirmed by observed data by \citet{R11}.
This relationship is theoretically derived from the timescale of the angular momentum loss via magnetic stellar wind \citep{T83, IT84}. 
In Fig.~\ref{fig:1}, this dependence of luminosity on the orbital period is represented by the dashed-dotted line.
In the same figure, persistent LMXBs containing less-evolved donors (filled triangles) and giant donors (filled squares) are plotted \citep[data are taken from ][]{R11, R12}.

The positions of NS-ULXs are shown in the same figure; the data used are listed in Table~\ref{table:1}. 
The two NS-ULXs located near the line extended into the long $P_{\rm{orb}}$ region. 
One (P-13) locates well below the line; its location seems to fit the same trend as that of LMXBs fed by giant donors. 
These locations that pretend the locations of LMXBs seems somehow odd, since the accretion time-scales are different between LMXBs and NS-ULXs. 
In low-mass binary systems, the mass transfer proceeds within the timescale of magnetic stellar wind \citep{T83, IT84}, and cause the $L_{\rm{X}} - P_{\rm{orb}}$ relation in Eq. ~(\ref{eq:LxPorb}). 
In contrast, NS-ULXs seem to have high-mass donors: $M_{\rm{D}} > 5 \rm{M}_{\odot}$ \citep{B14}.
These high-mass close binary systems would evolve within the thermal timescale of their donors \citep{KW96}, if the donor is less-evolved. 
Therefore, it is not appropriate to extend such a relation obtained for low-mass systems to the high-mass end, and
NS-ULXs are not necessarily located on the same $L_{\rm{X}} - P_{\rm{orb}}$ relation. 
The result that the locations of NS-ULXs show the similar trend with LMXBs might be caused by chance.

In the case of LMXBs, plenty of information can be obtained from the $L_{\rm{X}} - P_{\rm{orb}}$ relation \citep{IT84, R12}.
Hence, if we have the same kind of relationship appropriate for NS-ULXs, also it would be useful.
Especially for NS-ULXs, $L_{\rm{X}}$ and $P_{\rm{orb}}$ are simple and relatively easy parameters to be observed, though in general it is difficult to measure parameters of extra-galactic sources.
Thus, we try to establish the same kind of $L_{\rm{X}} - P_{\rm{orb}}$ relationship appropriate for NS-ULXs to investigate the nature of these systems.  

If the stellar radius achieves the Roche lobe radius before the termination of main sequence (TMS) or the subsequent Hertzsprung gap stage, the outer envelope of the star overflows its Roche lobe and falls onto the compact object via the first Lagrangian point. 
Here, to construct a new $L_{\rm{X}} - P_{\rm{orb}}$ relation appropriate for NS-ULXs, we need to compute the thermal time (and the consequent mass transfer rate) in the binary systems with certain orbital periods. 
This could be done by using an approximated stellar evolution track, given by \citet{HPT00}. 
Here, we compute evolutions of the stellar radii from the zero-age main sequence phase 
to the beginning of the giant branch (BGB) phase, by employing equations (1) to (30) in \citet{HPT00}.

The following procedure is used to construct an appropriate $L_{\rm{X}} - P_{\rm{orb}}$ relation for NS-ULXs. 
First, a pair of an orbital period and a donor mass is fixed. 
By using the canonical neutron stellar mass ($1.4 \rm{M}_{\odot}$), the Roche lobe radius around the donor is obtained by Eq. ~(\ref{eq:Rrl}).
By assuming that the donor star expands and fills its Roche lobe ($R_{\rm{donor}} = R_{\rm{RL}}$) at a certain stage of its evolution, the luminosity of the donor at this evolutionary stage is obtained. 
Consequently, the thermal time and the luminosity of the neutron star can be obtained. 
Repeating these procedures, we can obtain the $L_{\rm{X}} - P_{\rm{orb}}$ relation for high-mass systems. 
Examining the parameters of NS-ULXs of the $L_{\rm{X}} - P_{\rm{orb}}$ relation obtained, we would be able to draw a conclusion for the hidden nature of donors in NS-ULXs.

\begin{table*}
\centering
\caption{Parameters of three NS-ULXs: their name, maximum luminosity, orbital period, predicted donor mass, and mass accretion rate given in \citet{K17}. References: (a) \citet{B14}, (b) \citet{F15}, (c) \citet{T16}, (d) \citet{H17}, (e) \citet{I17b}, (f) this work, (g) \citet{M14}, (h) \citet{F16}, (i) \citet{I17a}.
}
\label{table:1}
\begin{tabular}{lcccccc} 
\hline
name & $L_{\rm{X, max}}$ [$\rm{erg \, s}^{-1}$] & $P_{\rm{orb}}$ [$\rm{days}$] & $M_{\rm{D}}$ [$\rm{M}_{\odot}$] & $\dot{M}_{\rm{NS}}$  [$\dot{\rm{M}}_{\rm{Edd}}$] & $ (8/3) \dot{M}_{\rm{NS}}$ [$\rm{g \, s}^{-1}$] & ref. \\
\hline
M82 X-2 & $2 \times 10^{40}$ & 2.5 & $5.2 - 10 $ & 36 & $1.54 \times 10^{20}$ & (a) (b) (c) (d)\\
NGC5907 ULX-1 & $1 \times 10^{41}$ & 5.3 & $6 - 10 $ & 91 & $3.88 \times 10^{20}$& (e) (f) \\
NGC7793 P-13 & $6 \times 10^{39}$ & 64  & $18 - 23 $ & 20 & $8.54 \times 10^{19}$ & (g) (h) (i) \\
\hline
\end{tabular}
\end{table*}


\section{Results}

In Fig.~\ref{fig:2}, our $L_{\rm{X}} - P_{\rm{orb}}$ relations are represented by solid curves. 
The curve labelled gTMSh denotes the relation assuming that the donor fills its Roche lobe, just after completing its hydrogen burning phase, and another curve labelled as gBGBh denotes this at the beginning of the giant branch phase. 
If the donor star is still in the main sequence, the plot on the $L_{\rm{X}} - P_{\rm{orb}}$ plane moves above the line labelled TMS.
However, if the donor has evolved beyond the giant branch phase, the plot on this plane no longer has any meaning, since the mass transfer from a giant donor does not proceed within the thermal time.
The region between the two lines corresponds to donors just passing the Hertzsprung gap.
In the same figure, we plot three NS-ULXs with filled symbols (circle, triangle and square for M82 X-2, ULX-1, and P13, respectively). 
The plotted parameters are summarized in Table~1.
In this figure, an important consequence can be seen; the donors in M82 X-2, ULX-1, and P-13 are in the Hertzsprung gap phase, near the termination of main sequence, and in the giant phase, respectively. 

Furthermore, masses of donors are given in this plot. 
The thin dashed lines in the figure connect the same mass models between TMS stage and BGB stage. 
The corresponding mass is shown near the lines in solar mass unit. 
In this figure, the plot of ULX M82 X-2 comes in the Hertzsprung gap region, between $4 \rm{M}_{\odot}$-line and $8 \rm{M}_{\odot}$-line. 
This location is consistent with the hitherto-indicated donor mass of this system, $5 \rm{M}_{\odot} < M < 10 \rm{M}_{\odot}$ \citep{B14,F15}.  
Additionally, it is indicated that the currently unknown donor mass of ULX-1 comes around $8 \rm{M}_{\odot}$-line.

In the regime of supper-Eddington accretion, the luminosity given by Eq.~(\ref{eq:lx}) may not be valid. 
Hence, here we try to examine our $L_{\rm{X}} - P_{\rm{orb}}$ relation from a different view. 
In \citet{K17}, they argued that in pulsating ULXs, accreted matter reaches the magnetospheric radius near the spherization radius, and then trapped by the magnetic field and falls onto the magnetic poles. 
Considering the effects of beaming, they estimated the accretion rates onto the neutron stars. 
Their derived mass accretion rates for three NS-ULXs are also shown in Table~1.  
Here, we compare this accretion rate with the mass transfer rate from the donor.

With this procedure, we show the corresponding $\dot{M} - P_{\rm{orb}}$ relation in Fig.~2 (right vertical axis) and plotted locations of three NS-ULXs with open symbols. 
Comparing with the previous plots, the positions of ULXs are slightly changed.  
Namely, the donor of ULX-1 could most probably at the TMS stage star in mass-range $4 - 8 \rm{M}_{\odot}$.
The estimated mass of M82 X-2 is almost unchanged. 
P-13 could enter still in the giant donor region.
Though the mass of each ULXs are somehow varied, the qualitative results are unchanged.  
From these results, we conclude that our theoretical examinations are qualitatively appropriate to estimate the nature of the donor star of NS-ULXs. 

\begin{figure}
	\includegraphics[width=\columnwidth]{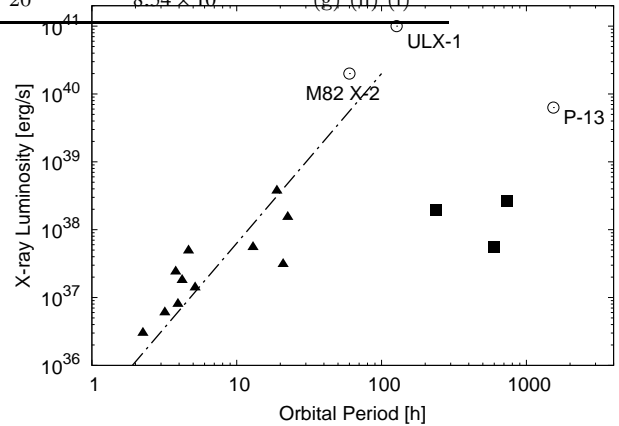}
\caption{
Position of NS-ULXs and persistent LMXBs on the $L_{\rm{X}} - P_{\rm{orb}}$ plane.
LMXBs with main sequence donors and giant donors are shown by filled triangles and filled squares, respectively. 
Data are taken from \citet{R11}. 
Three known NS-ULXs are also shown by open circles with labels of their name.
The dashed-dotted line denotes the $L_{\rm{X}} \propto P_{\rm{orb}}$ line obtained in \citep{IT84, R12}. 
}
    \label{fig:1}
\end{figure}

\begin{figure}
	\includegraphics[width=\columnwidth]{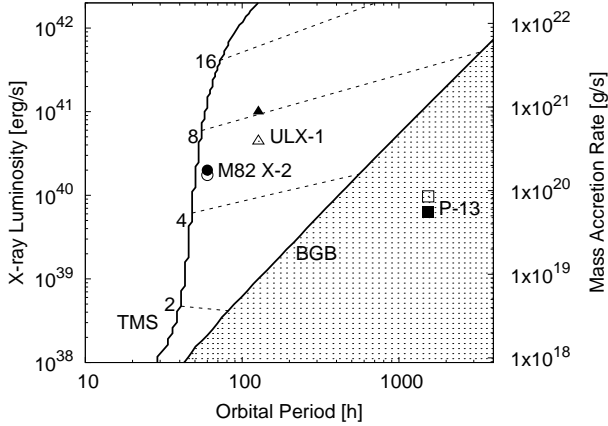}
\caption{
$L_{\rm{X}} - P_{\rm{orb}}$ relation (filled symbols) obtained by the approximated stellar evolution track. 
The upper solid curve in each panel denotes our theoretical $L_{\rm{X}} - P_{\rm{orb}}$ relation assuming that 
the donor fills its Roche lobe when it is at the terminal point of main sequence phase (TMS). 
The lower solid curve denotes that at the beginning of giant branch phase (BGB).
Three known NS-ULXs are shown by filled symbols with labels of their name \citep{B14,I17a,I17b,F16}.
Below the lower curve (shaded area), the mass transfer no longer proceeds within thermal time, 
and our theoretical relations do not have any sense. 
The corresponding $\dot{M} - P_{\rm{orb}}$ relation is also shown in the same figure.
The mass transfer rate given in \citet{K17} is shown by open symbols.
}
    \label{fig:2}
\end{figure}

The donor of P-13 is a highly evolved star and the timescale of its mass transfer is no longer of thermal time. 
Hence, the plotted position could not have any physical meaning. 
It is important, however, that the position of P-13 in this figure is below the region of less evolved systems. 
For LMXBs, the giant systems come systematically right-hand side of the $L_{\rm{X}} \propto P_{\rm{orb}}^{2.5}$ curve fitted for systems with less evolved donors (the dashed-dotted line in Fig.~\ref{fig:1}). 
P-13, where the donor is a giant, follows the same trend. 
This decrease of luminosity might be caused by the formation of a common envelope covering the X-ray source. 
Another possible cause is that for vastly expanded stars, the dynamical timescale ($\tau_{\rm{dyn}} \propto \rho^{-1/2}$) can increase. 

The orbital period of P-13 is inferred to be 64~days, which it is much longer than that of other similar systems \citep{M14,F16}. 
Hence, it is argued that this 64~day periodicity may be due to a super-orbital period. 
In this case, it is suggested that the true orbital period could be $7-9$ days, based on other similar systems showing super-orbital periodicities \citep{H17}. 
However, if we assume that $P_{\rm{orb}} = 8 \rm{d}$ and the luminosity is $10^{39} \rm{erg \, s}^{-1}$, the corresponding location deviates from the trend of LMXBs involving giant donors. 
If the similarity between LMXBs and NS-ULXs with giant donors were true, that might suggest that the 64~day period of P-13 is actually the orbital period of the binary system. 
Additionally, the luminosity of a super-giant which fulfills the Roche-lobe for a $P_{\rm{orb}} = 8 \rm{d}$ system is smaller than the observed value. 
Then, we consider that the orbital period of P-13 is most probably about 64~day \citep{M14}.

We note that, with a large mass ratio ($q = M_{\rm{D}} / M_{\rm{NS}}$), the Darwin instability might disturb the binary orbit \citep{C73}. 
The criterion for this type of instability is still debated, the critical mass ratio could be $q \simeq 12$ (for polytropic donor model with $n = 3$) \citep{E06}.  
Additionally, when the mass ratio is $\approx 1$, the orbit could become unstable. 
Hence, in the high-mass and low-mass end, we cannot apply the $L_{\rm{X}} - P_{\rm{orb}}$ relation, shown in Fig.~\ref{fig:2}. 
(For the recent discussion about the mass transfer process from massive stars, see, for example, \citet{Pi17}.)

Here, we have examined the position of three NS-ULXs on the recently constructed $L_{\rm{X}} - P_{\rm{orb}}$ plane. 
Using this new method, we can derive the currently unknown properties of the NS-ULXs, from only limited observational data, such as $L_{\rm{X}}$ and $P_{\rm{orb}}$. 
By using this procedure, we have studied the donor star in the NGC5907 ULX-1 system, and we have suggested that the donor can be a star with $6 - 10 \rm{M}_{\odot}$ at the terminal main sequence stage. 

\section{discussion}

The main aim of this study is to present the possibility of extracting information of donors in NS-ULX systems from limited observational properties, such as luminosities and orbital periods. 
Using this method, we can derive the currently unknown nature of the donor in the NGC5907 ULX-1; the donor could be a star with $6 - 10 \rm{M}_{\odot}$, after its hydrogen burning phase.

There are possibly a lot of ULXs, which have not been confirmed yet as NS-ULXs \citep{Pz17}, since binaries containing neutron stars are much more populous than black hole binaries \citep{BZ09}.
Additionally, the condition for a critically accreting neutron star to emit pulses, is very restricting, and only a tiny fraction of NS-ULXs would be observed as pulsars \citep{K17}. 
In general, it is difficult to investigate the donor stars in extragalactic binaries, even with large optical telescopes. 
Furthermore, when the nature of the compact companion (neutron star or black hole) is unknown, it is quite hard to confirm the masses and the evolutionary stages of donor stars \citep{G13}.
An independent method to derive information of donors only from easily observed properties, such as luminosities and orbital periods, would be useful. 
Recently, methods to extract the nature of X-ray sources of ULXs from their X-ray spectrum are proposed \citep{BGS17, Pz17}, and the sample of NS-ULXs will increase in the near future. 
Therefore, it is required to develop theoretical methods for analysis of NS-ULXs from many directions. 
As a first step in this direction, we consider the validity and applicability of our study.

\subsection{Validity of the $L_{\rm{X}} - P_{\rm{orb}}$ relation}

In this section, we consider the validity of our study, especially our $L_{\rm{X}} - P_{\rm{orb}}$ relation, shown in Fig.~\ref{fig:2}.
Using this figure, we can predict the mass and the evolutionary stage of donor stars. 
The $L_{\rm{X}} - P_{\rm{orb}}$ relation, however, contains several uncertainties.

One uncertainty is the mass of the neutron star. 
In this study, a canonical neutron star mass, $M_{\rm{NS}} = 1.4 \rm{M}_{\odot}$, was assumed.
However, it is suggested that the mass of neutron stars has a bimodal distribution; one peak appears around $1.4 \rm{M}_{\odot}$ and another appears at $1.8 \rm{M}_{\odot}$ \citep{A16}. 
Furthermore, in binary systems, especially in old populations like LMXBs, the mass of neutron stars tends to be greater than that of the isolated neutron stars. 
If a heavy neutron star, with $2 \rm{M}_{\odot}$ is considered, the curves in Fig.~\ref{fig:2} are moved {\it{upward}} $\sim 40 \%$.

Additionally, the computation of the X-ray luminosity from the mass transfer rate (Eq.~(\ref{eq:lx})) includes another uncertainty. 
This is due to the difference between the liberated potential energy of the accreted matter and the emitted X-ray energy. 
It is naturally assumed that the transferred matter from the donor cannot be accreted onto the neutron star smoothly and directly, as suggested by the transient nature of NS-ULXs \citep{DSD14, T16}. 
The transferred matter would be stopped in the accretion disc, due to magneto-centrifugal force, and fall into the accretion column, when the neutron star spins down \citep{DPS14}. 
Additionally, only a few 10\% of the potential energy of the accreted matter can be converted into X-rays at the surface of the neutron star \citep{ST83,W01}.
Therefore, we preferably need to add a factor of $\mathcal{O} (0.1)$ to the luminosity estimation. 
Consequently, the luminosity shown in Fig.~\ref{fig:2} can also vary one order of magnitude {\it{downward}}. 
(Note that this uncertainty can be balanced with the heavier mass of neutron star. )
We include this, as a parameter $\epsilon$ in Eq.~(\ref{eq:lx}), containing those uncertainties described above. 
Though the exact value of $\epsilon$ is unknown, we have adopted tentatively $\epsilon = 0.1$ here.

Actually, even if the $L_{\rm{X}} - P_{\rm{orb}}$ curve is shifted by one order of magnitude, the basic conclusions described above remain unchanged.  
That is, if we vertically shift the curves in Fig.~\ref{fig:2} by 50\%, the donor type of the M82 X-2 might remain in the Hertzsprung gap region. 
Also in this case, the position of ULX-1 is still in the Hertzsprung gap region. 
This is because the $L_{\rm{X}} - P_{\rm{orb}}$ curve for the TMS stage is steep, and the result is insensitive to vertical shifts. 
Additionally, the consequence of the donor of P-13 being a well-evolved giant could be the same. 
Although the donor mass of ULX-1 has been predicted to be $6 - 10 \rm{M}_{\odot}$ in Fig.~\ref{fig:2}, this result might be a tentative value. 
However, our result for the M82 X-2 shown in Fig.~\ref{fig:2} is quite consistent with the previous estimation of $5 \rm{M}_{\odot} < M < 10 \rm{M}_{\odot}$ \citep{B14,F15}, and this consistency strongly supports the validity of our estimation. 

In this study, we have assumed that a RLOF occurs when the stellar radius expands toward to the Roche lobe radius. 
In close binary systems that contain strong X-ray sources, however, this assumption should be examined. 
It is suggested that the RLOF might start before the donor expands to fill the Roche lobe due to strong X-ray irradiation \citep{ITF97, L16}.  
However, such an irradiation effect is typical for tight binary systems only, which have their orbital period decreased to $\sim 1 \rm{hr}$. 
Though our three NS-ULX systems have large X-ray luminosities, their orbital periods are much longer.
Hence, the irradiation effect might not important for these systems. 
In the future, however, if NS-ULXs with orbital periods less than $1 \rm{d}$ were found, the effects of X-ray irradiation should be considered.

In order to improve the accuracy of this method, it is desirable to perform calibrations by using pulsating HMXB systems with known donor masses. 
Unfortunately, however, HMXB systems fed via RLOF are quite rare in our galaxy, since the evolutionary time scale of relatively tight binaries ($P_{\rm{orb}} < 10 \rm{d}$), with Roche lobe filling massive donors is quite short \citep{P02}.
However, the derived property of the donor in M82 X-2 shows a good agreement with previously derived values (see Fig.~\ref{fig:2} and \citet{B14}).

\subsection{Donor mass of NGC5907 ULX-1}

In the first report of pulse detection from NGC5907 ULX-1, its donor mass has not been determined \citep{I17a}.
The authors only suggested three possibilities for the donor: (i) a less evolved massive ($> 10-15 \rm{M}_{\odot}$) star, (ii) a super-giant with a mass excess of $10 \rm{M}_{\odot}$, (iii) a giant with $>1 \rm{M}_{\odot}$ and (iii) a giant with $2-6 \rm{M}_{\odot}$ \citep{I17a}. 
In the above analysis, we have suggested that the donor mass of this system could be $\approx 10 \rm{M}_{\odot}$ (see Section 3). 
It seems consistent with the scenario (ii) in \citet{I17a}.
The scenarios (i) and (iv) may barely consistent with our result. 
Here, by taking the stellar evolution tracks into account, we attempt to further examine the properties of the donor.

\begin{figure}
\begin{center}
 	\includegraphics[width=\columnwidth]{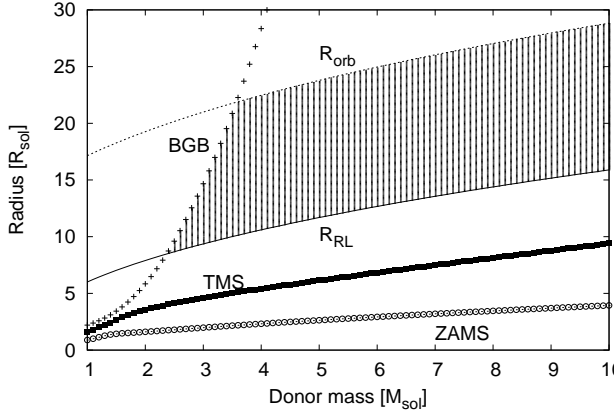}
   \caption{
The RLOF condition (shown by shaded area). 
We show the orbital radius given by Eq.~(\ref{eq:a}) (dashed curve) and the Roche lobe radius given by Eq.~(\ref{eq:Rrl}) (solid curve).
At the same time, stellar radii of the donor as functions of stellar mass are shown at ZAMS phase (open circles), TMS phase (filled squares), and BGB phase (crosses), respectively.
}
    \label{fig:3}
\end{center}
\end{figure}

By assuming a donor mass, the Roche lobe radius can be computed by eq.~(\ref{eq:Rrl}). 
Simultaneously, the stellar radius can be computed by using the approximated stellar evolution track, given by \citet{HPT00} for the assumed mass and its evolutionary stage. 
The obtained radius of the donor, as a function of the donor mass is shown in Fig.~\ref{fig:3}.
Donor radii at several evolutionary stages (zero age main sequence, the termination of the main sequence and the beginning of the giant branch) are also shown, and labelled as ZAMS, TMS, and BGB, respectively.
A primitive condition for the start of RLOF accretion is that the donor radius must exceed the Roche lobe radius. 
At the same time, the donor radius should be smaller than the orbital radius. 
The shaded region in Fig.~\ref{fig:3} shows where these conditions are satisfied.

In the case that the donor fills its Roche lobe, after entering the giant phase, a convective envelope sets in, and the mass transfer proceeds within the dynamical timescale \citep{E06}. 
Then, the accretion could no longer be stable and the common envelope would be formed. 
Since, the system in this case would be luminous in infra-red frequencies, rather than in X-rays \citep{I13}, the region above the line of BGB should be removed from the RLOF condition in Fig.~\ref{fig:3}. 

\begin{figure}
\begin{center}
	\includegraphics[width=\columnwidth]{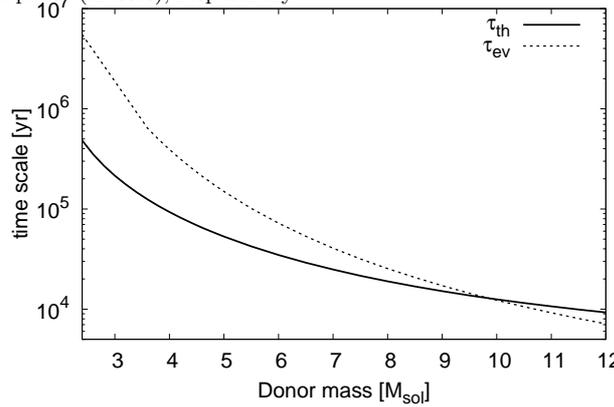}
    \caption{
Typical timescales.
The thermal timescale and evolution timescale are shown by solid and dashed curves, respectively.  
}
    \label{fig:4}
\end{center}
\end{figure}

According to Fig.~\ref{fig:3}, a system with a low-mass donor ($< 2.3 \rm{M}_{\odot}$) does not satisfy the RLOF condition, stated above. 
However, for $M_{\rm{D}}$ larger than $2.3 \rm{M}_{\odot}$, the star will be inside its Roche lobe during its main sequence lifetime, but may overflow it during its expansion, after the end of the main sequence (the Hertzsprung gap phase), giving an early case B type of mass transfer. 
This proceeds in the thermal timescale, as given in Eq.~(\ref{eq:tauth}), and in principle, could be sufficient to feed a ULX.
Up to this point, the upper limit of the donor mass is yet to be determined. 
In general, however, the timescale on which the star crosses the Hertzsprung gap is quite short, especially for massive stars. 
In Fig.~\ref{fig:4}, we show the relevant timescales for this system. 
If the donor mass is less than $2.3 \rm{M}_{\odot}$, the common envelope would be formed in the dynamical timescale,  
and it is remarkably shorter than both the thermal or the evolutionary time. 
However, if the donor mass is larger than $2.3 \rm{M}_{\odot}$, the mass transfer would proceed in the thermal timescale, especially for a compact binary with an intermediate-mass donor \citep{P02, W01}. 
The thermal time as a function of the donor mass is indicated by the solid line in the figure. 
The evolutionary time (defined by the duration that the radius of the donor enters the shaded region in Fig.~\ref{fig:3}) is shown by the dotted line. 
When the donor mass is larger than $10 \rm{M}_{\odot}$, the donor quickly expands during its Hertzsprung gap phase, and the evolutionary time dominates the mass transfer. 
With a massive donor, the observable lifetime of a ULX could only be several thousand years. 
Such a short timescale for the duration of accretion would limit a chance of observation. 
Hence, though scenarios (i) is also possible, the mass of donor should not be much larger than 10 Msol. 
Additionally, scenario (iv) is also consistent; but 2 Msol could be too small. 
As the conclusion, the systems with a moderately massive ($\approx 10 \rm{M}_{\odot}$) donor could be most probable candidate of ULX-1. 

The donor type of NS-ULXs are studied with population synthesis computations by \citet{W17}. 
They suggested that the largest population of the donors of NS-ULXs is red giants with $\sim 1 \rm{M}_{\odot}$, which belong typically old populations. 
However, it is not clear whether this result can be  directory applied to {\it pulsating} NS-ULXs in straightforward way.
Since X-ray pulsars typically have strong magnetic fields and this suggests that they are young systems. 
There are strict conditions on the magnetic field and the accretion rate for pulsating NS-ULXs \citep{K17}, and it may be difficult to satisfy these conditions for systems containing less-massive red-giant donors.
Additionally, we need to keep in mind that the parent galaxy of NS-ULXs are star forming galaxies \citep{B14, I17a}.
Though the majority of NS-ULX donors synthesized by \citet{W17} are less-massive, NS-ULXs that recognized due to X-ray pulsations might not be typical populations. 
Of course, our results do not negate scenarios (i) and (iii) in \citet{I17a} completely.
In scenario (i) or (iii), however, the possibility of observation could be very small. 
Dichotomy method between BH-ULXs and NS-ULX without pulsations could play an important role to verify the theoretical prediction given by population synthesis \citep{BGS17,W17}.

\section{Conclusions}

In this study, we have examined the position of the recently discovered three NS-ULXs on the $L_{\rm{X}} - P_{\rm{orb}}$ plane, which has been applied in the LMXB analysis. 
We have constructed a new $L_{\rm{X}} - P_{\rm{orb}}$ relation suitable for NS-ULXs, and derived the currently unknown properties of NS-ULXs. 
Especially, combining the stellar evolution tracks with the observed properties, we have drawn a well-founded conclusion about the donor star in the NGC5907 ULX-1 system: the donor could be a rate main-sequence or a Hertzsprung gap star, with $6 - 10 \rm{M}_{\odot}$.

\section*{Acknowledgements}

We thank the anonymous referee for helpful comments. 





\begin{thebibliography}{99}

\bibitem[\protect\citeauthoryear{Antoniadis et al.}{2016}]{A16}
Antoniadis, J., Tauris, T. M., Ozel, F., Barr, E., Champion, D. J. \& Freire, P. C. C.,
2016, preprint (arXiv:1605.01665)

\bibitem[\protect\citeauthoryear{Bachetti et al.}{2014}]{B14}
Bachetti, M. et al., 2014, Nature, 514, 202
\bibitem[\protect\citeauthoryear{Begelman}{2002}]{B02}
Begelman, M. C., 2014, ApJ, 568, 97
\bibitem[\protect\citeauthoryear{Belczynski \& Ziolkowski}{2009}]{BZ09}
Belczynski, K. \& Ziolkowski, J., 2009, ApJ, 707, 870
\bibitem[Bildsten et al.(1997)]{B97}
Bildsten, L. et al., 1997, ApJ Suppl., 113, 367
\bibitem[\protect\citeauthoryear{Burke et al.}{2017}]{BGS17}
Burke, M. J., Gilfanov, M. \& Sunyaev, R., 2017, MNRAS, 466, 194
  
\bibitem[\protect\citeauthoryear{Corbet}{1984}]{C84}
Corbet, R. H. D., 1984, A \& A, 141, 91
\bibitem[\protect\citeauthoryear{Counselman}{1973}]{C73}
Counselman III, C. C., 1973, ApJ, 180, 307

\bibitem[\protect\citeauthoryear{Dall'Osso et al.}{2015}]{DPS14}
Dall'Osso, S, Perna, R. \& Stella, L., 2015, MNRAS, 449, 2144
\bibitem[\protect\citeauthoryear{Doroshenko et al.}{2014}]{DSD14}
Doroshenko, V., Santangelo, A. \& Ducci, L., 2014, A \& A, 579, 22

\bibitem[\protect\citeauthoryear{Eggleton}{1983}]{E83}
Eggleton, P. P., 1983, ApJ, 268, 368
\bibitem[\protect\citeauthoryear{Eggleton}{2006}]{E06}
Eggleton, P. P., 2006, "Evolutionary processes in binary and multiple 
stars", Cambridge Univ. Press, Cambridge UK
\bibitem[\protect\citeauthoryear{Ek{\c s}i et al.}{2015}]{E14}
Ek{\c s}i, K. Y., Andac, I. C., Cikintoglu, S., Gencali, A. A., Gungor, 
C. \& Oztekin, F., 2015, MNRAS, 448, L40
  
\bibitem[\protect\citeauthoryear{Farrell et al.}{2009}]{F09}
Farrell, S. A., Webb, N. A., Barret, D., Godet, O. \& Rodrigues, J. M., 
2009, Nature, 460,73
\bibitem[\protect\citeauthoryear{Fragos et al.}{2015}]{F15}
Fragos, T., Linden, T., Kalogera, V. \& Sklias, P., 2015, ApJ, 802, 5
\bibitem[\protect\citeauthoryear{F{\"u}rst et al.}{2016}]{F16}
F{\"u}rst, F. et al, 2016, ApJ, 831, 14 

\bibitem[\protect\citeauthoryear{Gladstone et al.}{2013}]{G13}
Gladstone, J. C., Copperwheat, C., Heinke, C. O., Roberts, T. P., Cartwright, T. F., Levan, A. J. \& Goad, M. R., 
2013, ApJS, 206, 14

\bibitem[\protect\citeauthoryear{Hu et al.}{2017}]{H17}
Hu, C.-P., Li, K. L., Kong, A. K. H., Ng, C.-Y., Lin, C.-C., L., 2017, ApJ, 835, 9
\bibitem[\protect\citeauthoryear{Hurley et al. }{2000}]{HPT00}
Hurley, J. R., Pols, O. R. \& Tout, C. A., 2000, MNRAS, 315, 543

\bibitem[\protect\citeauthoryear{Iben \& Tutukov}{1984}]{IT84}
Iben, I., Jr. \& Tutukov, A. V., 1984, ApJ, 284, 719 
\bibitem[\protect\citeauthoryear{Iben et al.}{1997}]{ITF97}
Iben, I., Jr., Tutukov, A. V. \& Fedorova, A. V.\ 1997, ApJ, 486, 955 
\bibitem[\protect\citeauthoryear{Ivanova et al.}{2013}]{I13}
Ivanova, N., Justham, S., Avendano Nandez, J. L., \& Lombardi, J. C., 
2013, Science, 339, 433
\bibitem[\protect\citeauthoryear{Israel et al.}{2017a}]{I17a}
Israel, G. L., et al., 2017, Science, 355, 817
\bibitem[\protect\citeauthoryear{Israel et al.}{2017b}]{I17b}
Israel, G. L. et al., 2017, MNRAS, 466, 48

\bibitem[Kalogera \& Webbink(1996)]{KW96}
  Kalogera, V. \& Webbink, R. F., 1996, ApJ, 358, 301
\bibitem[\protect\citeauthoryear{Karino \& Miller}{2016}]{KM16}
Karino, S. \& Miller, J. C., 2016, MNRAS, 462, 3476
\bibitem[\protect\citeauthoryear{King et al.}{2017}]{K17}
King, A., Lasota, J.-P. \& Klu{\'z}niak, W., 2017, MNRAS, 468, 59

\bibitem[\protect\citeauthoryear{Lucy}{2017}]{L16}
Lucy, L. B., 2017, A \& A, 601, 75

\bibitem[\protect\citeauthoryear{Makishima et al.}{2000}]{M00}
Makishima, K. et al., 2000, ApJ, 535, 632 
\bibitem[\protect\citeauthoryear{Meyer \& Meyer-Hofmeister}{1979}]{MM79}
Meyer, F., \& Meyer-Hofmeister, E., 1979, A \& A, 78, 167 
\bibitem[\protect\citeauthoryear{Motch et al.}{2014}]{M14}
Motch, C., Pakull, M. W., Soria, R., Grise, F. \& Pietrynski, G., 2014, Nature, 514, 198 

\bibitem[\protect\citeauthoryear{Ohsuga \& Mineshige}{2007}]{OM07}
Ohsuga, K. \& Mineshige, S., 2007, ApJ, 736, 2
\bibitem[\protect\citeauthoryear{Ohsuga \& Mineshige}{2011}]{OM11}
Ohsuga, K. \& Mineshige, S., 2011, ApJ, 670, 1283
	
\bibitem[\protect\citeauthoryear{Pasham et al}{2014}]{P14}
Pasham, D. R., Strohmayer, T. E. \& Mushotzky, R. F., 2014, Nature, 513, 74
\bibitem[\protect\citeauthoryear{Pavlovskii et al.}{2017}]{Pi17}
Pavlovskii, K., Ivanova, N., Belczynski, K. \& Van., K. X., 2017, MNRAS, 465, 2092
\bibitem[\protect\citeauthoryear{Pintore et al.}{2017}]{Pz17}
Pintore, F., Zampieri, L., Stella, L., Wolter, A., Mereghetti, S. \& Israel, G. L., 2017, ApJ, 836, 113
\bibitem[\protect\citeauthoryear{Podsoadlowski et al.}{2002}]{P02}
Podsiadlowski, Ph., Rappaport, S. \& Pfahl, E. D., 2002, ApJ, 565, 1107

\bibitem[\protect\citeauthoryear{Revnivtsev et al.}{2011}]{R11}
Revnivtsev, M., Postnov, K., Kuranov, A. \& Ritter, H., 2011, A \& A, 526, A94
\bibitem[\protect\citeauthoryear{Revnivtsev et al.}{2012}]{R12}
Revnivtsev, M., Zolotukhin, I. Y. \& Meshcheryakov, A. V., 2012, MNRAS, 421, 2846
\bibitem[\protect\citeauthoryear{Ritter}{1976}]{R76}
Ritter, H., 1976, MNRAS, 175, 279 

\bibitem[Shapiro \& Teukolsky(1983)]{ST83}
  Shapiro, S. L. \& Teukolsky, S. A. \ 1983, "Black Holes, White Dwarfs, and Neutron Stars", John Wiley \& Sons, New York

\bibitem[\protect\citeauthoryear{Taam}{1983}]{T83}
Taam, R. E., 1983, ApJ, 268, 361
\bibitem[\protect\citeauthoryear{Takahashi \& Ohsuga}{2017}]{TO17}
Takahashi, H. \& Ohsuga, K., 2017, ApJ, 845, L9
\bibitem[\protect\citeauthoryear{Tsygankov al.}{2016}]{T16}
Tsygankov, S. S., Mushtukov, A. A., Suleimanov, V. F. \& Poutanen, J., 
2016, MNRAS, 457, 1101


\bibitem[Wellstein, Langer, Braun(2001)]{W01}
  Wellstein, S., Langer, N. \& Braun H., \ 2001, A \& A, 369, 939
\bibitem[\protect\citeauthoryear{Wiktorowicz et al.}{2017}]{W17}
Wiktorowicz, G., Sobolewska, M., Lasota, J.-P.  \& Belczynski, K., 2017, preprint 
(arXiv:1705.06155)



\bibitem[\protect\citeauthoryear{Zhu et al.}{2012}]{Z12}
Zhu, C., Lu, G., Wang, Z. \& Wang, N., 2012, PASP, 124, 195






\end{thebibliography}








\bsp	
\label{lastpage}
\end{document}